\documentstyle[pre,aps,multicol,epsf]{revtex}

\tighten

\begin{document}
\draft

%\preprint{12-19-95}
%\twocolumn[\hsize\textwidth\columnwidth\hsize\csname @twocolumnfalse\endcsname
\title{  Dynamical Breaking of Charge Neutrality   
in Intrinsic Josephson Junctions:  Common Origin 
for Microwave 
Resonant Absorptions and Multiple-branch Structures in the 
I-V Characterisitcs 
%Origin of Multiple-branch Structures in I-V Characterisitics 
%in Intrinsic Josephson Junctions 
%and I-V Characterisitcs in Layered High-Tc Superconductors
%Is It Intrinsic or Extrinsic  ? 
}

\author{$^{a}$ M. Machida }

\address{ 
Center for Promotion of Computational Science and Engineering,
\\ 
 Japan Atomic Energy Research Institute, 2-2-54 Nakameguro, Meguro-ku
Tokyo 153, Japan \\ 
$^{a}$ CREST, Japan Science and Technology Corporation (JST), Japan 
}

\author{ T.Koyama }
\address{ Institute for Materials Research, Tohoku University, Katahira
2-1-1, Aoba-ku, Sendai 980-77, Japan}

%\author{ A.Tanaka }
%\address{ Teikyo University of Science, Uenohara 2525, Yamanashi, Japan } 

\author{ M.Tachiki }
\address{
 National Research Institute for Metals, Sengen 1-2-1, Tsukuba, Ibaraki 305, Japan}

\date{\today}

\maketitle

\begin{abstract}

We demonstrate that both microwave resonant absorptions and
multiple-branch structures in the I-V characterisitcs observed in 
intrinsic Josephson junctions (IJJ's) are caused by dynamical 
breaking of charge neutrality (DBCN) inside the atomic-scale 
superconducting layer.
The Lagranginan for the time-dependent Lawrence-Doniach
model incorporating the effect of the DBCN is proposed, and 
the longitudinal collective Josephson 
plasma mode is proved to exist based on the Lagrangian. On the other hands, 
the branching behaviors in the I-V curves   
are almost completely reproduced by careful numerical simulations 
for the model equation derived from the Lagragian.

\end{abstract}
%\vskip2pc  

\begin{multicols}{2} \narrowtext

%\section{*}

Intrinsic Josephson effects (IJE's) in highly anisotropic 
high temperature superconductors (HTSC's) have attracted 
growing interests \cite{Tachiki1},\cite{Bulaevskii1}, and 
recent extensive experimental studies have discovered two 
remarkable phenomena for IJE's, i.e., the microwave resonant 
absorptions (MRA's) \cite{Matsuda1}
and the multiple-branch 
structures (MBS's) in I-V characterisitics along the c-axis \cite{Kleiner1}.
%%Although those phenomena are characteristic of the IJE's which
%%discriminate between the IJE's and the conventional Josephson effects, 
%%existing explanations for those are based on opposite interpretaions
%%for IJE's each other and the understanding for IJE's 
%%is now still controversial.    

In Bi-2212 single crystals(SC's), the MRA has been observed 
for both longitudinal 
and transverse microwave configurations in cavities, and the 
observed dispersion relations have clearly been distinguished from 
one another for those configurations \cite{Kakeya1}. The MRA in the
longitudinal configuration is caused by the excitaion of the 
longitudinal Josephson plasma mode propagating along the c-axis, and
%%is a collective charge fluctuation mode along the c-axis \cite{Tachiki2}.
%%The existence of such a wave-like charge mode strongly 
the existence of such the charge fluctuation mode 
indicates that the charge neutrality can be dynamically broken inside  
the superconducting layer under the application of an AC electric 
field along the c-axis.
Thus, a novel coupling between junctions due to the DBCN has been 
suggested and new collective dynamical behaviors 
have been expected to emerge \cite{Tachiki2},\cite{Koyama1}.  
%%therefore,  a novel coupling form between 
%%junctions due to the Coulomb interaction has been suggested 

On the other hands, the I-V curves along the c-axis in Bi-2212 SC's are 
composed of many branches with almost equal inter-branch spacging 
and the number of those is equal to the number of junctions \cite{Kleiner1},
\cite{Schlenga1}.
It has been widely believed that such structures manifests the 
independence of junctions inside IJJ's, that is, the
coupling between junctions is negligible, and then, each junction
behaves independently \cite{Schlenga1}. Thus, such a understanding    
for the I-V curves is irreconcilable with the above interpretation
for the MRA experiments, and the understanding for IJE's 
is now still controversial for the two experimental results.    

In this paper, we explain the DBCN inside the superconducting layer 
in IJJ's, and give a proper model Lagrangian for IJJ's. 
By using the Lagrangian we prove that the c-axis propagating 
longitudinal plasma mode exists, and show that the MBS 
in the I-V characteristics 
can be well reproduced by careful numerical simualtions 
for the model equation derived from the Lagrangian \cite{Koyama1},
\cite{Mac1}. 
Thus, we clarify that both the MRA and the MBS have 
the common origin, i.e., the DBCN. 

Now, let us explain about the DBCN in IJJ's.  
Figure.1(a) and (b), respectively, are a schematic figure   
for physical events followed by the electron tunneling   
in conventional stacked SIS junction (CSISJ) system 
and IJJ.  
From the figures, it is found that the charges   
are confined near the surface in the CSISJ system while 
the whole of the superconducting layer is charged in the IJJ,
 i.e., the DBCN occurs.
This is because the superconducting CuO$_2$ layer in IJJ's
is extremely thin ($\sim 3\AA $ ), i.e., the thickness is 
comparable to the charge screening length and therefore the charge 
screening becomes incomplete within the single superconducting layer. 
In CSISJ systems, the electric field generated at 
a junction site is completely screened out by sufficiently 
thick superconducting layers, and the coupling between 
different junction-sites is not formed.        
On the other hands, in IJJ's, %the emerged charge is screened 
%over some superconducting layers, and the single superconducting layer 
%itself is charged, i.e., the DBCN occurs.  
%Thus, 
the electric field generated inside a junction site 
affects the neighboring junction 
sites and therefore the coupling between junctions 
emerges \cite{Koyama1},\cite{Mac1}.
%Also, the Coulomb pontential screened incompletely inside 
%the superconducting layer affects charge dynamics in neighboring 
%superconducting layers.  

\begin{figure}
\centerline{\epsfxsize=9.0cm \epsfysize=5.0cm \epsfbox{Fig1.eps}}
\end{figure}
%\vskip -2.2cm
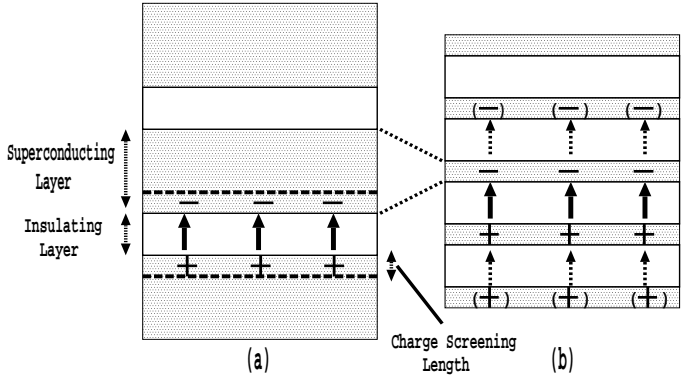
FIG.1~ Schematic views of (a) the conventional 
stacked SIS junction (CSISJ) and (b) the intrinsic Josephson junction.
The arrows indicate the direction of the
electron tunneling and the generated electric field.
In the CSISJ, the charges 
are screened within the range of the charge screening length 
 as indicated by the horizontal dotted lines, while in IJJ's 
the whole superconducting layer is charged due to 
the incomplete screening within the single superconducting layer. 
Consequently, the electric field generated in a junction site 
penetrates into the neighboring junction sites as the dotted arrows.  
\bigskip  

Here, let us give a proper Lagrangian, which describes
the above DBCN in IJJ's,
\begin{eqnarray}
L_{LD}= \sum_\ell %\Bigl\{{3e^{\ast 2}\Delta^2s\over 2m^\ast v_f^2}
\Bigl\{{ s  \over 8 \pi \mu^2  } 
\bigl(  A_0  (z_\ell, t )
+{\hbar\over e^\ast}\partial_t\theta_\ell\bigl)^2 \nonumber \\
- {\hbar\over e^{\ast}} j_c 
\Bigl( 1 - \cos P_{ \ell+1,\ell}\Bigl)
+ {\epsilon D\over 8\pi}E_{\ell+1,\ell}^2\Bigl\},  
\end{eqnarray}
where $s$ and $D$, respectively, are thickness of the
superconducting and insulating layer, $\mu$ is 
the charge screening length, $\epsilon$ is the dielectric constant, 
$E_{\ell+1,\ell}$ is the electric field ($\perp$ layers) between 
$(\ell+1)$th and $\ell$th superconducting layer,  
 $e^\ast=2e$, and $P_{\ell+1,\ell} (\equiv 
\theta_{\ell+1}(t)-\theta_\ell(t)- {e^\ast\over\hbar c} 
\int_{\ell D}^{(\ell+1)D} dz A_z(z,t) )$ is 
the gauge invariant phase difference. In eq.(1) 
the charge density of $\ell$-th superconducting layer, $\rho_\ell(t)$, is 
related to the scalar potential as 
$\rho_\ell(t)=-{1\over 4\pi\mu^2}(A_0(z_\ell)+{\hbar\over e^\ast}
\partial_t\theta_\ell)$ in the gauge-invariant form. This Lagrangian 
corresponds to a discrete version of the well-known time-dependent 
Ginzburg-Landau theory at $T=0$K \cite{Stephen1} \cite{Simanek} for 
stacked multi-Josephson-junction systems when
parameters $\mu$ and $j_c$ are taken as $\mu=\sqrt{m^\ast v_f^2/
(12\pi e^{\ast 2}\Delta^2)}$ and $j_c=\hbar e^\ast\Delta^2/(MD^2)$, 
where $v_f$ is the Fermi velocity, $\Delta$ is the amplitude of the order 
parameter at $T=0$K, and $m^\ast$ and $M$ are the effective masses parallel 
and perpendicular to the layers, respectively.
The first term in eq.(1) 
represents the interaction between the charge density and the gauge
invariant scalar 
potential.  In conventional theories for CSISJ 
systems, this term is dropped and the dynamics 
originating from only the 
2nd and 3rd terms in eq.(1) has so far been intensively studied 
\cite{Simanek},\cite{Ambegaokar1}. This 
is because the coefficient $s/\mu^2$ is considered to be very 
large in CSISJ systems
 and, thus, ${\cal L}$ is assumed to have a deep minimum at 
$\partial_t\theta_\ell=-(e^\ast/\hbar)A_0(z_\ell)$, which 
leads to the Josephson relation. Thus, the deviations 
around this minimum do not give low energy excitations 
and the Josephson relation always holds within the low 
energy fluctuations in terms of the phase dynamics
in such the CSISJ systems \cite{Ambegaokar1}. 
However, this is not the case in HTSC's, since the thickness, $s$, of the 
double superconducting CuO$_2$ planes is comparable with $\mu$. 
This means that 
the fluctuations arising from the first term in eq.(1) cannot be neglected 
in HTSC's, that is, the energy scale of the charge fluctuations inside the 
superconducting layers along the $c$-axis is comparable with that of the AC 
Josephson effect in HTSC's. This is the essential point which discriminates 
the IJE's from the AC Josephson effects in conventional serial 
Josephson-junction arrays.

Let us now prove that the system described by 
the Lagrangian (1) has the 
longitudinal plasma mode. 
To study the response to a longitudinal electric 
field along the $c$-axis, we derive the 
dielectric function in the present system. 
%%We notice that 
%%the Lagrangian (1) is reduced to the following form 
%%in the neutral case, 
%%\begin{equation}
%%{\cal L}=\sum_\ell\Bigl\{{s\over 8\pi\mu^2}
%%({\hbar\over e^\ast})^2(\partial_t\theta_\ell)^2
%%-{\hbar\over e^\ast}j_c\bigl[1-\cos (\theta_{\ell+1}-\theta_\ell)
%%\bigl]\Bigl\}.
%%\end{equation}
%%Note that $\mu e^\ast$ and $j_c/e^\ast$ are independent of the charge 
%%$e^\ast$ by definition. From eq.(2) it follows 
%%the Euler equation for $\theta_\ell$ in the linear approximation, 
%%as $\bigl[-\partial_t^2+v_B^2{1\over D^2}\Delta^{(2)}\bigl]
%%\theta_\ell=0$, 
%%where the velocity $v_B$ is given by $v_B=\sqrt{{8\pi\mu^2\over s}
%%({e^\ast\over\hbar})^2{\hbar\over 2e^\ast}j_c}D$ and the 2nd-rank 
%%difference operator $\Delta^{(2)}$ is defined as $\Delta^{(2)}
%%\theta_\ell\equiv \theta_{\ell+1}-2\theta_\ell+\theta_{\ell-1}$. 
%%The Euler equation describes the massless phason mode with the dispersion 
%%relation, $\omega_{\rm GB}(q)={v_B\over D}\sqrt{2(1-\cos q(s+D)}$, which 
%%is the Goldstone mode propagating in the direction perpendicular 
%%to the junctions in the neutral super-fluid system. 
As is well known, in charged systems the Goldstone mode is absorbed 
into the longitudinal 
gauge field and the gauge field becomes massive ( the Anderson-Higgs 
mechanism ). In order to describe this situation correctly,  %in the present 
%charged junction-arrays, 
it is convenient to utilize the phason gauge 
proposed by Matsumoto and Umezawa \cite{Matsumoto1}, in which the 
gauge condition is given by 
$
\partial_tA_0(z_\ell)+{v_B^2\over c}{A_{\ell+1,\ell}^z-A_{\ell+1,\ell}^z
\over D}=0
$, where $A_{\ell+1,\ell}^z\equiv \int_{\ell D}^{(\ell+1)D}dz A_z(z)/D$, 
and $v_B$ is the velocity of the phason mode and is given by 
$\sqrt{{8\pi\mu^2\over s}({e^\ast\over\hbar})^2{\hbar\over 2e^\ast}j_c}D$. 
Here, note that the neutral version of Lagrangian (1), in which 
dynamics of the gauge field $A_0 (z_\ell)$ and $A_{\ell+1,\ell}^z$ are 
dropped, gives the Euler equation for $\theta_\ell$ 
in the linear approximation 
as $\bigl[-\partial_t^2+v_B^2{1\over D^2}\Delta^{(2)}\bigl]
\theta_\ell=0$, where  $\Delta^{(2)}$ is defined as $\Delta^{(2)}
\theta_\ell\equiv \theta_{\ell+1}-2\theta_\ell+\theta_{\ell-1}$.
For the gauge transformation, 
$A_{\ell+1,\ell}^z\rightarrow A_{\ell+1,\ell}^z+(\hbar c/e^\ast D)
(\chi_{\ell+1}-\chi_\ell)$, $A_0(z_\ell)\rightarrow A_0(z_\ell)
-(\hbar/e^\ast)\partial_t\chi_\ell$, the phason gauge condition imposes the 
equation for $\chi_\ell$, $
\bigl[-\partial_t^2+v_B^2{1\over D^2}\Delta^{(2)}\bigl]
\chi_\ell=0$, which is the same as the 
Euler equation for $\theta_{\ell}$ in the neutral case. 
This result 
implies that the phason mode can be eliminated by the gauge transformation 
in the phason gauge, and the Lagrangian (1) is rewritten in the phason 
gauge as follows, 
\begin{eqnarray}
{\cal L}=\sum_\ell\Bigl\{{s\over 8\pi\mu^2}\bigl(A_0(z_\ell)\bigl)^2
-{\hbar\over e^\ast}j_c\bigl(1-\cos {e^\ast D\over \hbar c}
A_{\ell+1,\ell}^z\bigl) \nonumber \\
+{\epsilon D\over 8\pi}E_{\ell+1,\ell}^2\Bigl\}.
\end{eqnarray}
Variation of $A_0$ for eq.(2) yields 
one of the Maxwell equations, 
\begin{equation}
E_{\ell+1,\ell}-E_{\ell,\ell-1}=-{s\over\epsilon\mu^2}A_0(z_\ell) .
\end{equation}
From the relation (3), $E_{\ell+1,\ell}=-\partial_tA_{\ell+1,\ell}^z/c-
[A_0(z_{\ell+1})-A_0(z_\ell)]/D$, and the phason
gauge condition, by eliminating the vector 
potential, we also have another relation between the electric field 
and the scalar potential, 
\begin{equation}
E_{\ell+1,\ell}-E_{\ell,\ell-1}={D\over v_B^2}\bigl(\partial_t^2
-{v_B^2\over D^2}\Delta^{(2)}\bigl)A_0(z_\ell).
\end{equation}
Then, from eqs.(3) and (4) one finds the dielectric function for the 
electric field perpendicular to the junctions,  
\begin{equation}
\epsilon(\omega,k_z) 
=1-{\omega_{\rm pl}^2\over \omega^2-2\alpha\omega_{\rm pl}^2
[1-\cos k_z(s+D)]}.
\end{equation}
In deriving eq.(5) we used the relations, 
${s\over\epsilon\mu^2}{v_B^2\over D}={4\pi e^\ast D\over\epsilon\hbar}
={c^2\over\epsilon\lambda_c^2}=\omega_{\rm pl}^2$ and 
${2v_B^2\over D^2}={8\pi\mu^2\over s}{e^\ast\over\hbar}j_c=2\alpha
\omega_{\rm pl}^2$. As seen in eq.(5), the dielectric function has 
zero points at 
$\omega(k_z)=\omega_{\rm pl}\sqrt{1+2\alpha[1-\cos k_z(s+D)]}$, and 
the longitudinal plasma mode propagating along the $c$-axis is found to 
exist. Here, it should be noted that when 
the first term in the Lagrangian (1) is dropped 
the obtained Josephson plasma mode has no dispersion.    
%Thus, the system described by the Lagrangian (1) has 
%the longitudinal plasma mode.

Next, let us examine dynamics of the Euler-Lagrange equation 
for the gauge invariant phase difference $P_{\ell+1,\ell}(t) $ 
derived from the Lagrangian (1) in order to investigate 
the resistive states in IJJ's. 
The equation is given by two Maxwell equations derived from 
the variation of $A_0$ and $A_z$ for the Lagrangian and the modified 
Josephson relation obtained by using the gauge invariant charge
density expression, $
\partial_tP_{\ell+1,\ell}(t)={e^\ast\over\hbar}V_{\ell+1,\ell}(t)
-{4\pi\mu^2e^\ast\over\hbar}(\rho_{\ell+1}(t)-\rho_\ell(t))
$, as follows, 
\begin{eqnarray}
& &{1\over\omega_{\rm pl}^2}\partial_t^2P_{\ell+1,\ell}(t) 
   + \sin P_{\ell+1,\ell}(t) + {\beta\over\omega_{\rm pl}}
\partial_tP_{\ell+1,\ell}(t) = {I\over j_c} \nonumber \\
&-&\alpha\bigl(
\sin P_{\ell+2,\ell+1}(t)-2\sin P_{\ell+1,\ell}(t)
+\sin P_{\ell,\ell-1}(t)\bigl), 
\end{eqnarray}
where the transport current $I$ and the dissipative (quasi-particle)
current are introduced and those satisfy the current 
conservation relation, ${4\pi\over c}\Bigl(j_c \sin\theta_{\ell+1,\ell} 
+ \sigma  E_{\ell+1,\ell}
\Bigl) + {\epsilon\over c}\partial_tE_{\ell+1,\ell} = I$.
Also, $\omega_{pl}={c\over\sqrt{\epsilon}\lambda_c}$ and   
$\beta={4\pi\sigma\lambda_c\over\sqrt{\epsilon} c} 
(\equiv \frac{1}{\sqrt{\beta_c} })$, where $\beta_c$ is 
the McCumber parameter, and 
$\alpha (\equiv  \frac{\epsilon {\mu}^2 }{s D}) $ gives
the coupling between junctions.
It is clearly found that 
the parameter $\alpha$ can not be neglected only in 
IJJ's due to its definition. This model equation
has been phenomenologically given by two of us (T.K. and M.T. )
\cite{Koyama1}, and afterwards, Ch.Preis et al. 
confirmed that the same equation can be derived in their leading order 
\cite{Preis1}.
Here, note that if the first term of $L_{LD}$ is fixed to be zero,
the model equation becomes the RCSJ model \cite{Ambegaokar1}, i.e.,  
the model equation for independent junctions. 
Also, it should be noted that when $I$ and $\beta$ are assumed to be zero 
the linearized eq. (6) has a plane wave solution with the same dispersion 
relation as the relation derived from eq.(5) 
%that derived from zero points of the dielectric function 
\cite{Koyama1},\cite{Mac1}.
%%$\omega ( k_z )= \omega_{pl} \sqrt{1+2\alpha 
%%( 1-\cos k_z (s+D) ) } $ for small  $ P_{\ell+1,\ell}$ 
%%in the case of $\beta=0$ and $I=0$.
%%This excitation mode may be identified with
%%the c-axis propagating longitudinal Josephson plasma mode, which   
%%has been observed in the MRA of Bi-2212 SC's. 
In order to obtain %calculate 
%I-V curves
%and reproduce 
the MBS, we numerically solve eq.(6) 
for ten identical junctions
under the periodic boundary condition.
The values of parameters in eq.(6)  
$\alpha$ and $\beta$, respectively, are
chosen as 0.1 and 0.20, supposing Bi-2212 case.
Although in our previous paper \cite{Mac1} $\alpha$ was taken as 2.27,  
the present value employed was proved to be reasonable  
in Bi-2212 \cite{Preis1}. 

Figure 2(I) is a hysterisis loop in I-V curve for eq.(6). 
In obtaing this curve, the current I is increased first up 
to a value above the critical current $j_c$ and then decreased 
to zero along the arrows in Fig.2(I).  
In the current increasing process,  we find a jump at $j_c$, where
all junctions synchronously switch into same resistive states.  
In Fig.2(II), the time developments of the normalized Josephson 
currents
 $sin P_{\ell+1,\ell}$ at the point B are plotted for all junction
sites.
As seen in this figure, all junctions show a synchronous whirling motion.
%Such the motion appears in the wide range from points A to B on the 
%I-V curve. 
In decreasing the current from the point B, we find several
step-like structures as seen in Fig.2(I).
Figure 2(III) and (IV) show the time developments of $sin P_{\ell+1,\ell}$
of all junction sites at the point C and D, respectively. 
It is found that some junctins are in the resistive state, in which 
$sin P_{\ell+1,\ell}$ periodically varies from -1 to 1, while 
the others are in the superconducting state, in which it 
%$sin P_{\ell+1,\ell}$
shows tiny oscillatory motions. 
By monitoring the time evolution of $sin P_{\ell+1,\ell}$ in
the region around the point C and D as Fig.2(III) and (IV), one may see that 
%the step-like structures arise from the changes in 
%the number of the resistive junctions. 
in decreasing the current %some of the 
%resistive junctions jump into the superconducting state and 
the number of the resistive junctions decreases together with the step-like 
structures as seen in Fig.2(III) and (IV). 
Such the behaviors are never observed in the case without 
the coupling, i.e., $\alpha=0$ in eq.(6), which shows only a switching 
behavior into the superconducting state of all resistive 
junctions at a constant current value.
\begin{figure}
\centerline{\epsfxsize=7.5cm \epsfysize=6.0cm \epsfbox{Fig2r.eps}}
\end{figure}
%\vskip -2.2cm
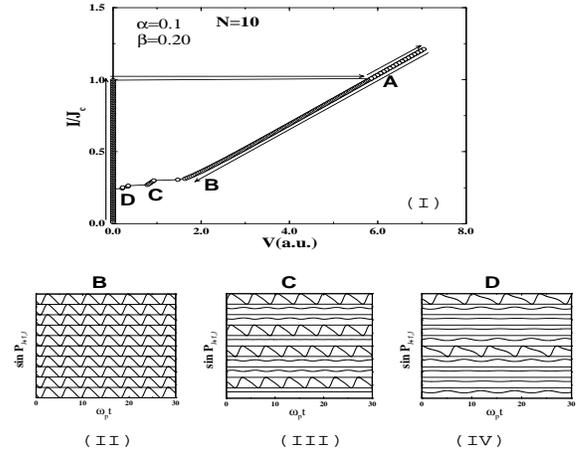
FIG.2~(I) The I-V hysterisis curve obtained by the current 
increasing and decreasing along the arrows.
The time developments of $sin P_{\ell+1,\ell}$ of all junctions
from $\ell = 0$ to $\ell = 9$ at (II) the point B, (III) 
the point C, and (IV) the point D in the I-V hysterisis curve (I).    
\bigskip  

Let us now study the region around the point C and D
%, in which the step-like structures are observed, 
in more details.
In Fig.3(a), we present the I-V curve enlarged in 
the region around C and D.
In this region, we can get different I-V curves
when intervals $\delta I$ of the decrease in the current value
is changed. 
In the Fig.3(a), two cases with different $\delta I$ are plotted.
Those behaviors imply that several bifurcation 
points are distributed in a very narrow region
for the controlling parameter $I$ of eq.(6).
This is a contrast to the case of $\alpha=0$. %, in which   
%all junctions jump into the superconducting states at 
%the same current value. 
Thus, in order to obtain all possible branches in  
trajectories of eq.(6) in the current decreasing process, we perform 
the calculations
many times, changing the decreasing current interval $\delta I$. 
Then, when the current is reincreased from 
suitable points of the obtained branches
 along the arrows as seen in 
Fig.3(a), we can obtain 
several I-V branches, which finally lead to the MBS as shown in Fig.3(b).
The simulated MBS is composed of equidistant
branches and the number of resistive branches (=10) is equal to
the number of junctions (=10).
However, in the present simulation, it should be noted
that we introduce Nyquist noise to get 
the first and ninth branches. 
%%Without the noise the dynamical transition to those two branches
%%could not be observed, %%though careful current 
%%decreasing simulations were performed 
But, 
after the transitions by the introduction of 
the noise, those two branches could be stably traced 
without the noise
in the current reincreasing. 

\begin{figure}
\centerline{\epsfxsize=6.0cm \epsfysize=5.0cm \epsfbox{Fig3abr.eps}}
\end{figure}
%\vskip -2.2cm
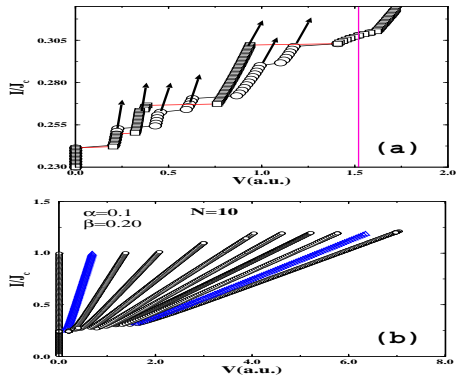
FIG.3~(a) The enlarged view of the transition region in I-V curve.
The circles and square points show different I-V curves obtained by 
changing the interval of the decrease in the current value. 
(b) The MBS in the I-V curve  
obtained by the current reincreasing process along the arrows 
as indicated in (a). 
\bigskip

\noindent Thus, all the resistive branches  
can be understood to correspond to stable trajectories in the 
phase space of eq.(6), and 
the MBS as observed in Bi-2212 SC's can be reproduced by 
eq.(6) without 
inhomogeneities of junction parameters.
In addition, we mention that if a free boundary condition, in which
the top and bottom junctions are distinguished from 
internal ones, is employed, we can obtain all the branches
without any noise.
In the actual experiments for the MBS, the current 
decreasing and increasing are perfomed many times.
The present simulations also require
a number of careful current decreasing and 
increasing process.     

Finally, let us consider why eq.(6) can give 
MBS's as observed in Bi-2212. The question resolves itself into the 
following two 
points. The first one is why several step-like transitions occur 
in decreasing the current as seen in Fig.3(a) 
 and the second one is why
those branches can keep their stablilities 
in the current reincreasing process as seen in Fig.3(b). 
In this paper, we focus on only the first point.
This is because  Takeno et al . 
have already touched the second point in a context for 
stabilities of localized whirling (rotating) motions 
in the different coupled rotator models with the coupling 
given by trigonometric functions \cite{Takeno1}   and they proved that   
the bounded character of the trigonometric function is essential 
for the existence of the localized rotating motions \cite{Takeno1}. 
As shown before, we observe 
that those step-like jumps in the current decreasing %(see Fig.3(a))
result from the stepwise transitions of resistive junctions into 
the superconducting state as seen in 
Fig.2(III) and (IV). 
%can be classified by the number of junctions in the 
%resistive state, so that the transition between nearest-neighbor branches 
%in the current decreasing process may be caused by the appearance of a 
%superconducting junction ( 
This implies that although all junctions are identical
each junction can switch into the superconducting state 
at a different current value $I/j_c$.
These behaviors 
closely resemble those of the independent RCSJ model under the presence of   
inhomogenious distributions in junction parameters.
% in the case of 
%non-zero $\alpha$. 
%superconducting junctions can coexist with neighboring 
%resistive junctions in the present model. 
We easily notice from the right hand side of eq.(6) that the total driving 
force ( current )
acting on a junction can be classfied into two components, which  
are the normalized external current $I/j_c$ and  
the normalized Josephson current in neighboring 
junctions,  
$I_{nbj}=  -\alpha\bigl( \sin P_{\ell+2,\ell+1}(t)
+\sin P_{\ell,\ell-1}(t)\bigl)$, where the term $2\alpha sin P_{\ell+1,\ell}$ 
is transposed to the left hand side.  
%work as a driving current in addition to the external current. 
Thus, it is clearly found that the total 
driving force acting on each junction depends on the 
dynamical state of neighboring junctions. 
Here, in order to demonstrate it, let us suppose that 
$\ell$-th junction is the resistive state and consider two simple cases, 
in which 
both of the neighboring $(\ell\pm 1)$-th 
junctions are in the superconducting state and both 
are in the resistive one. 
%%In such the cases, due to differences in the total driving currents
%%acting on $\ell$-th junction, it is found that the 
%%the external current values leading to
%%the superconducting transtion for two cases are different.   
In the former case, since the neighboring junctions show 
only the tiny oscillations around the stable points,  
$I_{nbj}$ is very small ( $I_{nbj}\sim 0$ ), and the total driving 
current almost equal to the external current $I/j_c$.  
%Consequently, $\ell$-th junction 
%shows the superconducting transition at almost the same value as that 
%of the RCSJ model, i.e., eq.(6) with $\alpha=0$.
On the other hands, in the latter case, 
%as long as three ($\ell$-th and $(\ell\pm 1)$th) junctions 
%do not show the in-phase motions, %( see above the stability analysis of 
%the in-phase whirling state ), 
the minimum of the instanteneous driving current is given 
by $I/j_c-2\alpha$, and 
it can sufficiently become smaller than that of the former case.
This clearly implies that in the current decreasing process 
the superconducting transition of $\ell$-th junction
in the latter case occurs prior to the former case. 
%In fact, the simulation results
Thus, it is found that the switching current value of each junction 
is dependent on dynamical states of neighboring junctions and therefore
the superconducting transition points in eq.(6) are distributed 
in the range of the controlling parameter $I$. 
%its influence is spread over the 
%neighboring junction sites as seen in Fig.(1). 
From those considerations, it is found that  
the origin of the dynamical transitions between branches 
is the coupling between junctions due to the charge incomplete screening 
within the single superconducting layer, i.e., the  
DBCN  inside the superconducting layer. % as well as the MRA.

In conclusion, we show that the DBCN inside the 
superconducting layer is essential for atomic-scale 
stacked Josephson junction systems as IJJ's, and give the proper 
model Lagragian for IJJ's describing the DBCN.  
We clarify that the Lagragian involves the dynamics of 
the longitudinal propagating Josephson 
plasma responsible for the MRA observed in the longitudinal configuration
while the Euler-Lagrange equation under the transport and 
the disspative current can reproduce the MBS.  
%
%is given, and  
%The equation gives the longitudinal collective 
%oscillation mode propagating along c-axis in the superconducting state, while
%it is shown that careful numerical simulations for the resistive states can 
%well reproduce MBS observed in experiments.      
%Also, the instability of the in-phase motion 
%is theoretically predicted and dynamical step like
%branching behaviors are explained by the existence of the coupling in the 
%model equation. 

%+++++++++++ SHIORI +++++++++++++++++++++++++++

The authors thank C.Helm and H.Matsumoto for illuminating discussions 
and one of us (M.M.) also thank A.Tanaka for useful discussions and
K. Asai and H. Kaburaki for their supports 
on the computer simulations in JAERI.

%\end{document}

%%%%%%%%%%%%%%%%%%%%%%%%%%%%%%%%%%%%%%%%%%%%%%%%%%%%%%%%%%%%%%%%%%%%%%%%
%%

\end{multicols}\widetext

\end{document}